\documentclass[graybox]{svmult}

\usepackage{natbib}
\usepackage{mathptmx}       
\usepackage{helvet}         
\usepackage{courier}        
\usepackage{type1cm}        
\usepackage{makeidx}         
\usepackage{graphicx}        
\usepackage{multicol}        
\usepackage[bottom]{footmisc}

\makeindex             

\begin{document}
\newcommand{\meandnu} {\langle\Delta\nu\rangle}

\title*{Frequency dependence of $\Delta \nu$ of solar-like oscillators investigated: Influence of HeII ionization zone}
\titlerunning{Frequency dependence of $\Delta \nu$ of solar-like oscillators investigated}
\author{S. Hekker, Sarbani Basu,  Y. Elsworth, W.J. Chaplin}
\institute{S. Hekker \at Astronomical Institute "Anton Pannekoek", University of Amsterdam, Science Park 904, 1098 HX Amsterdam, the Netherlands, \email{S.Hekker@uva.nl}
\and Sarbani Basu \at Department of Astronomy, Yale University, P.O. Box 208101, New Haven CT 06520-8101, USA
\and Y. Elsworth \at School of Physics and Astronomy, University of Birmingham, Edgbaston, Birmingham B15 2TT, UK 
\and W.J. Chaplin \at School of Physics and Astronomy, University of Birmingham, Edgbaston, Birmingham B15 2TT, UK}
%
%
\maketitle

\abstract*{Oscillations in solar-like oscillators tend to follow an approximately regular pattern in which oscillation modes of a certain degree and consecutive order appear at regular intervals in frequency, i.e. the so-called large frequency separation. This is true to first order approximation for acoustic modes. However, to a second order approximation it is evident that the large frequency separation changes as a function of frequency. This frequency dependence has been seen in the Sun and in other main-sequence stars. However, from observations of giant stars, this effect seemed to be less pronounced. 
We investigate the difference in frequency dependence of the large frequency separation between main-sequence and giant stars using YREC evolutionary models. In particular, we investigate the influence of the position and shape of the Helium second (HeII) ionization zone in the first adiabatic exponent as a function of evolution and stellar parameters, such as mass, metallicity, mixing length parameters, the inclusion of diffusion and different prescriptions of the atmosphere.
From this investigation we find that for less evolved stars the shallow location of the HeII zone enhances the dependence of the large frequency separation on frequency, compared to more evolved stars in which the HeII zone is located much deeper. Furthermore, the shape of the HeII zone in the first adiabatic exponent can be directly linked to the amplitude of the variation in the large frequency separation.}

\abstract{Oscillations in solar-like oscillators tend to follow an approximately regular pattern in which oscillation modes of a certain degree and consecutive order appear at regular intervals in frequency, i.e. the so-called large frequency separation. This is true to first order approximation for acoustic modes. However, to a second order approximation it is evident that the large frequency separation changes as a function of frequency. This frequency dependence has been seen in the Sun and in other main-sequence stars. However, from observations of giant stars, this effect seemed to be less pronounced.
We investigate the difference in frequency dependence of the large frequency separation between main-sequence and giant stars using YREC evolutionary models.}

\section{Introduction}
\label{sec:1}

Stellar oscillations are an important means to study the internal structure of stars. Stars with turbulent outer layers, such as low-mass main-sequence stars, subgiants and red-giant stars, can stochastically excite oscillations in their turbulent atmospheres, so-called solar-like oscillations. These oscillations generally follow a regular pattern in frequency described to reasonable approximation by the asymptotic relation derived by \citet{tassoul1980}:
\begin{equation}
\nu_{n,\ell} \approx \Delta\nu(n+\ell/2+\epsilon)-\ell(\ell+1)D_0,
\label{tassoul}
\end{equation}
with $\nu_{n,\ell}$ the frequency of an oscillation mode with radial order $n$ and degree $\ell$ and $\Delta \nu$ the large frequency separation between modes of the same degree and consecutive orders. $D_0$ is most sensitive to deeper layers in the star and $\epsilon$ to the surface layers.

The large frequency separation can be determined for many stars in a straightforward manner. It can be shown that this is directly proportional to the square root of the mean density of the star \citep{ulrich1986,kjeldsen1995}. Hence, this parameter plays an important role in studies concerning the internal structures of stars with solar-like oscillations. To a first order approximation the large frequency separation is constant over the observed frequency range. However, small deviations can occur due to for instance acoustic glitches, i.e., sudden internal property changes, but also due to other more slowly varying underlying variations, such as changing conditions close to the core. 

In contrast to the clearly observed frequency dependence of $\Delta\nu$ in the Sun and main-sequence stars \citep[e.g.][Fig. 17]{mathur2010}, first results from the \textit{Kepler} mission \citep{borucki2010} showed that for red giants the frequency dependence of $\Delta\nu$ is much less \citep{hekker2011comp}. We investigate this difference in sensitivity of $\Delta\nu$ to the frequency range for main-sequence stars and red giants using models constructed using YREC, the Yale stellar evolution code \citep{demarque2008}. In these models we use OPAL opacities \citep{iglesias1996} supplemented with low temperature ($\log T < 4.1$) opacities of \citet{ferguson2005} and the OPAL equation of state \citep{rogers2002}. All nuclear reaction rates are obtained from \citet{adelberger1998}, except for that of the $^{14}N(p,\gamma)^{15}O$ reaction, for which we use the rate of \citet{formicola2004}. 

For a more details about this study and a complete description of the different sequences of models used in this investigation and , we refer to \citet{hekker2011dnu} and references therein.

\section{Results}
\label{sec:2}

\begin{figure*}
\begin{minipage}{0.5\linewidth}
\includegraphics[scale=.3]{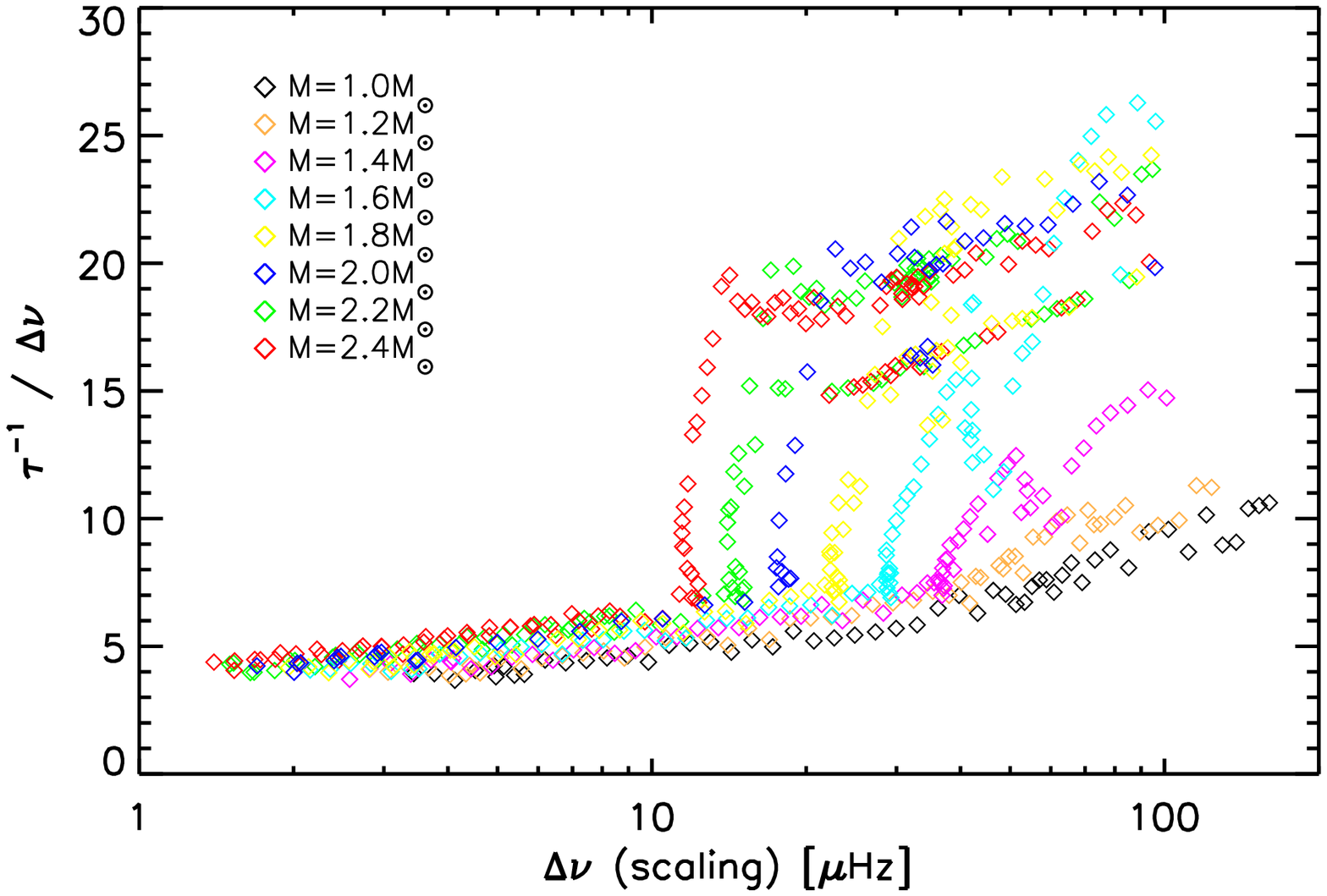}
 \end{minipage}
 \hfill
\begin{minipage}{0.5\linewidth}
\includegraphics[scale=.3]{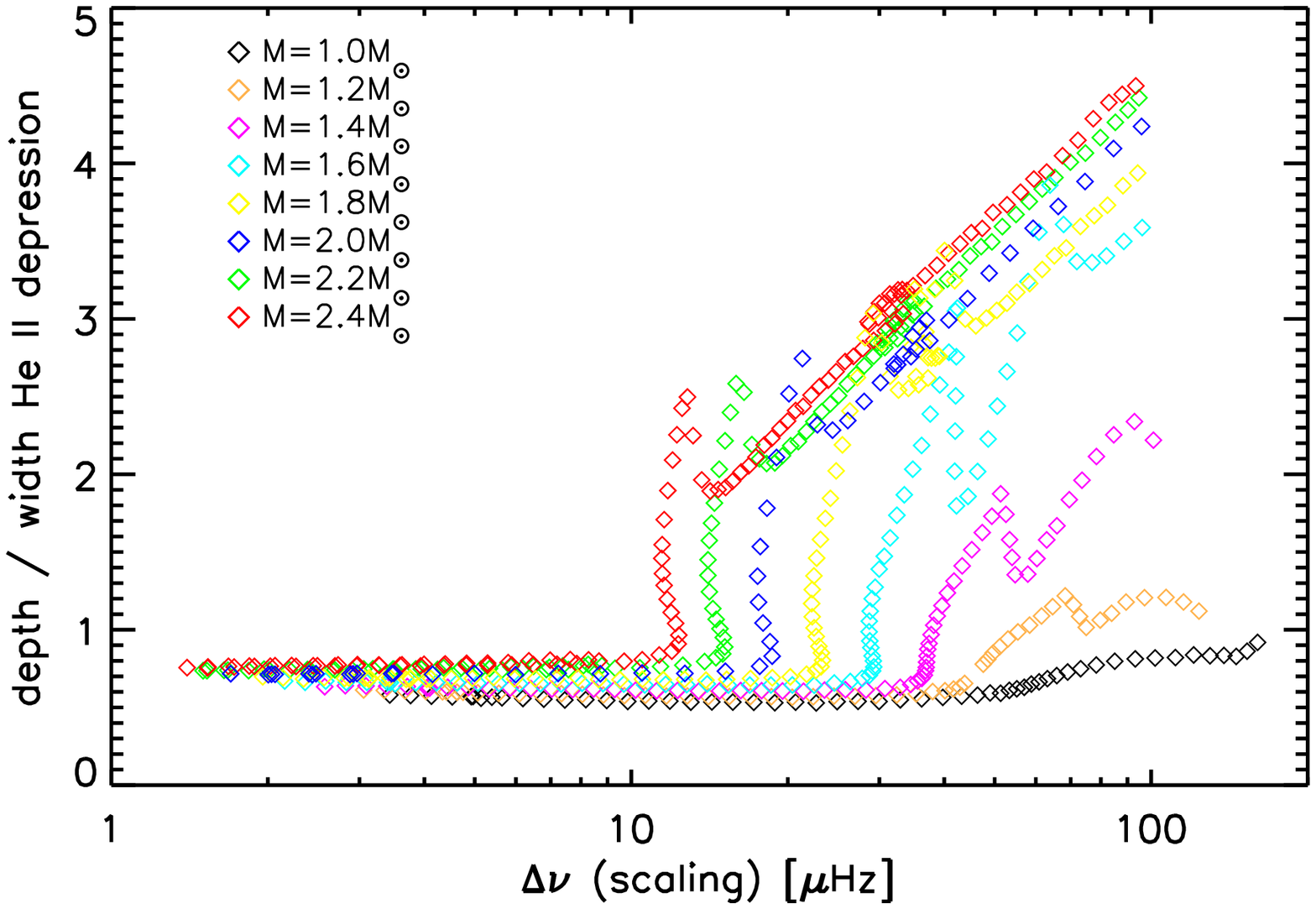}
\end{minipage}
\caption{Left: Period of the frequency modulation expressed in units of $\Delta \nu$ as a function of $\Delta \nu$. Right: The depth of the He II depression in $\Gamma_1$ divided by the width of the depression expressed in acoustic radius as a function of $\Delta \nu$.}
\label{fig:1}       
\end{figure*}

The observed difference in frequency dependence between main-sequence and red-giant stars could indeed be reproduced using radial oscillations computed for the YREC models. Hence,  we can investigate the cause of this difference using these models. 

First, we checked whether the variation of $\Delta\nu$ as a function of frequency is merely a trend, which can be resembled by a straight line with a certain slope, or due to faster variations with an amplitude dominant over the slope of a straight line. For more evolved stars the slope appears to be clearly dominant, which points to slowly varying underlying variations, such as changing conditions close to the core to be the dominant source of the frequency dependence. For less evolved stars the amplitude of the variation of $\Delta\nu$ around a straight line is dominant, which implies a dominant contribution of acoustic glitches to the frequency dependence. Interestingly, this amplitude appears to increase with increasing stellar mass and decreasing metallicity.

To investigate this further, we study the position and the shape of the depression in the first adiabatic exponent ($\Gamma_1~=~(\partial \ln p / \partial \ln \rho)_s$, with $p$ pressure, $\rho$ density and $s$ specific entropy) caused by the HeII ionization zone. This glitch modulates the frequencies in a sinusoidal manner with a so-called `period' inversely proportional to the acoustic depth ($\tau$) at which the glitch is located, with 
\begin{equation}
\tau=\int_r^R \frac{dr}{c},
\end{equation}
in which $r$ indicates the radius of the glitch, $R$ the radius of the star and $c$ sound speed. For more evolved stars the HeII zone is located deeper in the star causing a modulation in the frequencies, i.e. dependence of $\Delta\nu$ on frequency, with a `period' much shorter (of the order of 5$\Delta\nu$) than for less evolved stars, for which the `period' is much longer, up to 25$\Delta\nu$ (see left panel of Fig.~\ref{fig:1}). This confirms that for evolved stars the smooth stellar structure changes are the dominant dominant effect causing the dependence of $\Delta\nu$ on frequency, while for less evolved stars the effect from the HeII acoustic glitch is dominant.	\newline
\newline
We also investigated the influence of the shape of the depression in $\Gamma_1$ due to the HeII zone on the frequency dependence of $\Delta\nu$. We parametrize the shape of the HeII glitch by the ratio of the depth of the depression in $\Gamma_1$, i.e. its strength, to the width (in units of acoustic radius) of the depression. The narrower the width for a given depth the larger the effect the HeII depression has on the frequencies.
We find that the changes in the shape of the HeII depression in $\Gamma_1$ as a function of $\Delta \nu$ (see right panel of Fig.~\ref{fig:1}) follow the same pattern as the amplitude of the variation in $\Delta\nu$ around the trend. Therefore, we conclude that the shape of the HeII depression in $\Gamma_1$ is a significant cause of the difference in the amplitudes of the variation in $\Delta \nu$. Also, the increasing amplitude of the variation in $\Delta \nu$ as a function of mass and decreasing metallicity for less evolved stars is consistently present in the shape of the He II zone. However, there must be additional effects that play a role, because we only find qualitative agreement between the shape of the HeII depression in $\Gamma_1$ and the amplitude of the variation in $\Delta\nu$. An additional effect could be the influence of the base of the convection zone, which induces a small modulation with a period that is of the same order as the period of the HeII zone, but lower amplitude.

\begin{acknowledgement}
SH acknowledges financial support from the Netherlands Organisation for Scientific Research. SB acknowledges NSF grant AST-1105930. YE and WJC acknowledge financial support of UK STFC.
\end{acknowledgement}


\begin{thebibliography}{99.}
\bibitem[\protect\citeauthoryear{{Adelberger}, {Austin}, {Bahcall},
  {Balantekin}, {Bogaert}, {Brown}, {Buchmann}, {Cecil}, {Champagne}, {de
  Braeckeleer}, {Duba}, {Elliott}, {Freedman} \& {et al.}}{{Adelberger}
  et~al.}{1998}]{adelberger1998}
{Adelberger} E.~G.,  {Austin} S.~M.,  {Bahcall} J.~N.,  {Balantekin} A.~B.,
  {Bogaert} G.,  {Brown} L.~S.,  {Buchmann} L.,  {Cecil} F.~E.,  {Champagne}
  A.~E.,  {de Braeckeleer} L.,  {Duba} C.~A.,  {Elliott} S.~R.,  {Freedman}
  S.~J.,    {et al.} 1998, Reviews of Modern Physics, 70, 1265

\bibitem[\protect\citeauthoryear{{Borucki}, {Koch}, {Basri}, {Batalha},
  {Brown}, {Caldwell}, {Caldwell}, {Christensen-Dalsgaard}, {Cochran},
  {DeVore}, {Dunham}, {Dupree}, {Gautier}, {Geary}, {Gilliland}, {Gould} \& {et
  al.}}{{Borucki} et~al.}{2010}]{borucki2010}
{Borucki} W.~J.,  {Koch} D.,  {Basri} G.,  {Batalha} N.,  {Brown} T.,
  {Caldwell} D.,  {Caldwell} J.,  {Christensen-Dalsgaard} J.,  {Cochran} W.~D.,
   {DeVore} E.,  {Dunham} E.~W.,  {Dupree} A.~K.,  {Gautier} T.~N.,  {Geary}
  J.~C.,  {Gilliland} R.,  {Gould} A.,    {et al.} 2010, Science, 327, 977

\bibitem[\protect\citeauthoryear{{Brown}, {Gilliland}, {Noyes} \&
  {Ramsey}}{{Brown} et~al.}{1991}]{brown1991}
{Brown} T.~M.,  {Gilliland} R.~L.,  {Noyes} R.~W.,    {Ramsey} L.~W.,  1991,
  ApJ, 368, 599

\bibitem[\protect\citeauthoryear{{Demarque}, {Guenther}, {Li}, {Mazumdar} \&
  {Straka}}{{Demarque} et~al.}{2008}]{demarque2008}
{Demarque} P.,  {Guenther} D.~B.,  {Li} L.~H.,  {Mazumdar} A.,    {Straka}
  C.~W.,  2008, AP\&SS, 316, 31

\bibitem[\protect\citeauthoryear{{Ferguson}, {Alexander}, {Allard}, {Barman},
  {Bodnarik}, {Hauschildt}, {Heffner-Wong} \& {Tamanai}}{{Ferguson}
  et~al.}{2005}]{ferguson2005}
{Ferguson} J.~W.,  {Alexander} D.~R.,  {Allard} F.,  {Barman} T.,  {Bodnarik}
  J.~G.,  {Hauschildt} P.~H.,  {Heffner-Wong} A.,    {Tamanai} A.,  2005, ApJ,
  623, 585

\bibitem[\protect\citeauthoryear{{Formicola}, {Imbriani}, {Costantini},
  {Angulo}, {Bemmerer}, {Bonetti}, {Broggini}, {Corvisiero}, {Cruz},
  {Descouvemont}, {F{\"u}l{\"o}p}, {Gervino}, {Guglielmetti} \& {et
  al.}}{{Formicola} et~al.}{2004}]{formicola2004}
{Formicola} A.,  {Imbriani} G.,  {Costantini} H.,  {Angulo} C.,  {Bemmerer} D.,
   {Bonetti} R.,  {Broggini} C.,  {Corvisiero} P.,  {Cruz} J.,  {Descouvemont}
  P.,  {F{\"u}l{\"o}p} Z.,  {Gervino} G.,  {Guglielmetti} A.,    {et al.} 2004,
  Physics Letters B, 591, 61

\bibitem[\protect\citeauthoryear{{Hekker}, {Elsworth}, {De Ridder}, {Mosser},
  {Garc{\'{\i}}a}, {Kallinger}, {Mathur}, {Huber}, {Buzasi}, {Preston}, {Hale},
  {Ballot}, {Chaplin}, {R{\'e}gulo} \& {et al.}}{{Hekker}
  et~al.}{2011}]{hekker2011comp}
{Hekker} S.,  {Elsworth} Y.,  {De Ridder} J.,  {Mosser} B.,  {Garc{\'{\i}}a}
  R.~A.,  {Kallinger} T.,  {Mathur} S.,  {Huber} D.,  {Buzasi} D.~L.,
  {Preston} H.~L.,  {Hale} S.~J.,  {Ballot} J.,  {Chaplin} W.~J.,  {R{\'e}gulo}
  C.,    {et al.} 2011, A\&A, 525, A131
  
  \bibitem[\protect\citeauthoryear{{Hekker} {Basu} {Elsworth} \& {Chaplin}}{{Hekker}
  {et al.}}{2011a}]{hekker2011dnu}
{Hekker} S.,  {Basu} S., {Elworth} Y., {Chaplin} W.J., 2011a, arXiv: 1109.2595

\bibitem[\protect\citeauthoryear{{Iglesias} \& {Rogers}}{{Iglesias} \&
  {Rogers}}{1996}]{iglesias1996}
{Iglesias} C.~A.,  {Rogers} F.~J.,  1996, ApJ, 464, 943

\bibitem[\protect\citeauthoryear{{Kjeldsen} \& {Bedding}}{{Kjeldsen} \&
  {Bedding}}{1995}]{kjeldsen1995}
{Kjeldsen} H.,  {Bedding} T.~R.,  1995, A\&A, 293, 87

\bibitem[\protect\citeauthoryear{{Mathur}, {Garc{\'i}a},{R{\'e}gulo}, {Creevey}, {Ballot}, {Salabert}, {Arentoft}, {Quirion}, {Chaplin}, {Kjeldsen}}
{{Mathur} {et  al.}}{2010}]{mathur2010}
{Mathur} S., {Garc{\'i}a} R.A., {R{\'e}gulo} C., {Creevey} O.L., {Ballot} J., {Salabert} D., {Arentoft} T., {Quirion} P.-O., {Chaplin} W.J., {Kjeldsen} H., 2010, A\&A, 511, A46


\bibitem[\protect\citeauthoryear{{Rogers} \& {Nayfonov}}{{Rogers} \&
  {Nayfonov}}{2002}]{rogers2002}
{Rogers} F.~J.,  {Nayfonov} A.,  2002, ApJ, 576, 1064

\bibitem[\protect\citeauthoryear{{Tassoul}}{{Tassoul}}{1980}]{tassoul1980}
{Tassoul} M.,  1980, ApJS, 43, 469

\bibitem[\protect\citeauthoryear{{Ulrich}}{{Ulrich}}{1986}]{ulrich1986}
{Ulrich} R.~K.,  1986, ApJ, 306, L37

\end{thebibliography}
\end{document}